\documentclass[a4paper,10pt,onecolumn,noshowpacs,prd]{revtex4}
\usepackage{graphicx}
\usepackage{amsmath}
\usepackage{amssymb}
\usepackage{amsfonts}
\usepackage{bm}
\usepackage{textcomp}
\usepackage[usenames,dvipsnames]{xcolor}
\usepackage[colorlinks, linkcolor={RedOrange},citecolor={Purple}]{hyperref}

\def\be{\begin{equation}}
\def\ee{\end{equation}}
\def\bea{\begin{eqnarray}}
\def\eea{\end{eqnarray}}

\newcommand{\f}[2]{\frac{#1}{#2}}
\begin{document}
\title{Energy conditions in mimetic-$f(R)$ gravity}
\author{Zahra Haghani$^1$}
\email{z.haghani@du.ac.ir}
\author{Maryam Shiravand$^1$}
\email{m.shiravand@std.du.ac.ir}
\author{Shahab Shahidi$^1$}
\email{s.shahidi@du.ac.ir}
\affiliation{$^1$ School of Physics, Damghan University, Damghan, 41167-36716, 
Iran.}
\date{\today}

\begin{abstract}
The energy conditions of mimetic-$f(R)$ gravity theory is analyzed. We will obtain the parameter space of the theory in some special forms of $f(R)$ in which the self-acceleration is allowed. In this sense, the parameter space is obtained in a way that it violates the strong energy condition while satisfying the weak, null and dominant energy conditions. We will also consider the condition that the Dolgov-Kawasaki instability is avoided. This condition will be further imposed in the parameter space of the theory. We will show that the parameter space of the mimetic-$f(R)$ gravity is larger than $f(R)$ gravity theory.
\end{abstract}
\maketitle
\section{Introduction}
Instead of searching for an action which is invariant under the conformal transformations, one can define a new metric tensor which is conformally invariant by itself. Writing any gravity action with this new metric tensor will then results in a conformally invariant gravity theory. This interesting idea was recently proposed by Chamseddin and Mukhanov \cite{CM} where they have defined a new metric tensor, made of a dynamical metric tensor and a scalar field as
\be
g_{\mu\nu}=-{\hat g}_{\mu\nu}\partial_\alpha\phi\partial_\beta\phi {\hat g}^{\alpha\beta}.\label{moa1}
\ee
The metric $g_{\mu\nu}$ is invariant under the Weyl transformation on $\hat{g}_{\mu\nu}$ as
\be
\hat{g}_{\mu\nu}\rightarrow\Omega^2(x)\hat{g}_{\mu\nu}.\label{2}
\ee
Writing the usual Einstein-Hilbert action with the above new metric leads to a gravitational theory with an extra degree of freedom which behaves like a dust  \cite{CM}. So, one has a new pressure-less matter fluid with geometric origin which can play the role of cold dark matter sector of the universe. The theory is then  dubbed ``mimetic dark matter". An immediate consequence of definition \eqref{moa1} is that the scalar field should satisfy the relation
\be\label{cons}
g^{\mu\nu}\partial_\mu \phi \partial_\nu \phi=-1.
\ee
This means that the gradient of the scalar field should be time-like vector field everywhere. One can add the above equation into the action as a constraint through Lagrange multiplier. It turns out that both actions yield the same field equations \cite{Golovnev}. Many works have been done in the context of mimetic dark matter theory, including the Hamiltonian analysis \cite{hami}, cosmological perturbations \cite{pert}, and other generalizations of the theory \cite{gb}.

The constraint equation actually breaks the Lorentz invariance of the action by defining a preferred time-like direction in the space-time. Such a theory has been considered vastly in the literature, known as the ``Einstein-aether" theory \cite{tedi}. In this theory a time-like vector field, known as the aether vector, is introduced to the theory through a Lagrange multiplier. Many works has been done in the context of Einstein-aether theory, including cosmology \cite{EAcosmo} and black hole solutions \cite{EABH}. One of the interesting facts about the Einstein-aether theory is that it can be considered as a low energy limit of the Horava-Lifshitz gravity if the aether vector is hypersurface orthogonal \cite{HL}. In order to make the aether vector hypersurface orthogonal, one can define the aether as the gradient of a scalar field $T$ as
$$V_\mu=\f{\nabla_\mu T}{\sqrt{g^{\alpha\beta}\nabla_\alpha T\nabla_\beta T}}.$$
This changes the Einstein-aether theory to a gravitational theory with an extra scalar degree of freedom with a timelike gradient. Assuming that the gradient of the scalar field $T$ has a unit norm, makes the Einstein-aether theory and the mimetic dark matter theory very similar. The main difference between these two theories is that the Einstein-aether theory has some self-interaction terms for the aether vector.
 The first attempt to add some kinetic terms (other than the canonical kinetic term which is already present in the theory) to the action of mimetic dark matter theory was done in \cite{SEA} by writing the scalar version of the Einstein-aether theory using the substitution $V_\mu=\nabla_\mu\phi$. The theory can then be considered as a scalar Einstein-aether gravity. Unfortunately, the resulting theory is equivalent to the projectable Horava-Lifshitz gravity which is known to have ghost and strong coupling problems. This problem can simply be solved by adding to the theory a potential term which breaks the shift symmetry on the scalar field $\phi$ \cite{SEA}. However, one can add some other, i.e. healthy, self-interactions to the mimetic gravity action which makes the theory different from the projectable Horava-Lifshitz gravity, and in the mean time keeps the shift symmetry on the scalar field. In \cite{LVG}, the authors have added galileon self-interactions to the mimetic action making sure that the theory does not suffer from the Ostrogradski instability. The resulting theory can then explain the late-time accelerated expansion even in the dust dominated universe. In \cite{vik}, the authors have considered the term $\gamma(\phi)(\Box\phi)^2$ to the mimetic action which simultaneously has higher derivative terms and breaks the scalar shift symmetry. It is interesting to note that the Einstein-Hilbert action plus $\gamma(\Box\phi)^2$ term with constant $\gamma$ in the context of mimetic theory can be written as
 $$S=\int dt d^3x N\sqrt{-q} \big(~^3R+K_{ij}K^{ij}+(\gamma-1)K^2\big),$$
 where $K_{ij}$ is the extrinsic curvature, $K$ is its trace and $q_{\mu\nu}$ is the metric on the spacial hypersurface. This ADM form of the action shows that the following action breaks the Lorentz invariant and produces the canonical kinetic term of the Horava-Lifshitz gravity \cite{jaki}.
 One can also add some potential term for the scalar field to the action in order to explain the effect of dark energy in this model \cite{vik1}.

Beside adding higher order interaction terms to the action, one can promote the geometry to obtain an $f(R)$ gravity theory. The first interesting consequence of the $f(R)$ theory with the effective metric \eqref{moa1} is that the universe accelerate exponentially \cite{noji}. In \cite{fRMDM}, the authors have considered the dynamical system analysis of such model and showed that the universe has a stable accelerated expanding mode.

The mimetic-$f(R)$ theory can be considered as a gravitational theory coupled to a geometrical dark matter field which corresponds to the scalar field $\phi$. The modification of geometry is then responsible for the late-time accelerated expansion of the universe. By rewriting the theory as a Brans-Dicke theory with $\omega=0$, where $\omega$‌ is the Brans-Dicke coupling constant, the expansion of the universe will be encoded to a dynamical scalar field. In this sense the model can be considered as a gravitational theory coupled to two scalar degrees of freedom which are responsible for both dark sectors of our universe.

Despite the fact that the $f(R)$ gravity theories are very interesting and useful, they may suffer from instabilities. In 2003, Dolgov and Kawasaki \cite{DK} showed that the higher derivative theories may have an instability dropping the theory to a strong curvature regime in a very short time. One can obtain a constraint on the form of $f(R)$ to avoid such an instability \cite{insfR}. It is very interesting that the Palatini-$f(R)$ gravity does not have such an instability \cite{PfR}. In \cite{noji1} the authors found an example of $f(R)$ gravity which passes the Dolgov-Kawasaki instability. The Dolgov-Kawasaki instability has been studied in the various versions of higher derivative gravities, including theories with non-minimal coupling between matter and geometry \cite{fRT}. In the case of mimetic-$f(R)$ gravity, the results is the same as the standard $f(R)$ gravity \cite{DK}, due to the fact that the dark matter sector can not be responsible for the Dolgov-Kawasaki instability.

The main purpose of the present work is to consider the energy conditions of the mimetic-$f(R)$ theory. In general relativity, in order to have a viable description of the matter distribution, one has to impose certain conditions, known as the energy conditions, which are obtained from Raychaudhuri equation \cite{econgr}. These conditions basically ensure that the gravitational force is attractive and the energy density is positive definite. The Strong, Weak, Null and Dominant energy conditions are the most fundamental energy conditions, which are the key role for cosmology, black hole physics and thermodynamics. In cosmology, in order to have an accelerated expanding solution, one should have an extra degree of freedom which play the role of repulsive gravity. This means that the strong energy condition should be violated for this new degree of freedom, while is satisfied for ordinary baryonic matter. In this paper, we will analyze the energy conditions of the mimetic-$f(R)$ gravity theory, and obtain the parameter space of the theory corresponding with the violation of the strong energy condition. We will also take into account the criteria imposed by the Dolgov-Kawasaki instability. In this sense, three different/important choices of the function $f(R)$, i.e. the power law, the exponential case and also the mixed case will be considered in details. The energy conditions was investigated for many modified gravity theories, including $f(R)$ gravity, Brans Dicke, teleparallel gravity \cite{many}, $f(G)$ gravity \cite{fg}, $f(R,L_m)$ gravity \cite{frtenergy} and $f(R,T,R^{\mu\nu}T_{\mu\nu})$ gravity theories \cite{frtrt}.

The present paper is organized as follows: In the next section we will obtain the field equations of the model and construct its effective energy momentum tensor. The Dolgov-Kawasaki instability analysis will be discussed in section \ref{secDK}. In section \ref{secEC} we will obtain the energy conditions and consider some special cases for the function $f(R)$ in section \ref{some}. In section \ref{secDIS} we will summarize the results and make some discussions.
\section{The Model}
Let us consider the $f(R)$ gravity action in the mimetic theory
\be
\label{miFRaction}
S=\int d^4 x \sqrt{-g\left({\hat g}_{\mu\nu}, \phi \right)}
\left( f\left(R\left({\hat g}_{\mu\nu}, \phi \right)\right)
+ \mathcal{L}_{\mathrm{m}}\right)\, ,
\ee
where $\mathcal{L}_\mathrm{m}$ is the matter Lagrangian and the physical metric $g_{\mu\nu}$ is related to the auxiliary metric $\hat{g}_{\mu\nu}$ and the scalar field $\phi$ via equation \eqref{moa1} which yields the constraint equation \eqref{cons}.
The field equations can be obtained by varying the action with respect to the metric $\hat{g}_{\mu\nu}$ and the scalar field $\phi$ with the result
\begin{align} \label{eqg}
 \frac{1}{2} f g_{\mu\nu}
&-f^\prime R_{\mu\nu}
+ \nabla_\mu
\nabla_\nu
f'- g_{\mu\nu}
\Box 
f^\prime + \frac{1}{2} T_{\mu\nu} 
 + \partial_\mu \phi \partial_\nu \phi
\left( 2 f- R
f^\prime- 3 \Box
f^\prime+ \frac{1}{2} T \right)=0,
\end{align}
and
\be\label{eqphi}
\nabla^\mu
\left[\partial_\mu \phi \left( 2 f- R f^\prime - 3 \Box
f^\prime + \frac{1}{2} T \right) 
\right]=0,
\ee
where \textit{prime} denotes derivative with respect to the argument, and $T$ is the trace of energy-momentum tensor.

By taking the covariant derivative of the metric field equation \eqref{eqg} one can obtain
\be
\nabla^\mu T_{\mu\nu}=0,
\ee
where we have used the scalar field equation \eqref{eqphi} and
$
\nabla^\nu \phi\nabla^\mu \nabla_\nu \phi=0,
$
obtained by taking the covariant derivative of \eqref{cons}. So, the ordinary matter energy-momentum tensor is independently conserved in this theory. This suggests that the massive test particles in this theory move along the geodesics, and no extra forces is present here \cite{fRT,FRT}.

From equation \eqref{eqphi}, one obtains
\be\label{phi2}
\sqrt{-g}\left(2f -R f^\prime -3\Box f^\prime+\f{1}{2}T\right)\partial_\mu \phi=C_\mu,
\ee
where $C_\mu$ is an integration constant. It is clear from equation \eqref{phi2} that the vector density $C_\mu$ is either zero or time-like.
One can see from equations \eqref{eqg} and \eqref{eqphi} that in the case of vanishing $C_\mu$ the theory reduces to the standard $f(R)$ gravity. In this paper we will consider both non-zero and zero values of $C_\mu$.

Substituting equation \eqref{phi2} into \eqref{eqg}, one obtains
\begin{align}\label{Ein}
G_{\mu\nu}=\f{T_{\mu\nu}}{2f^\prime}+\left(\f{f}{2f^\prime}-\f{1}{2}R\right)g_{\mu\nu}+\f{1}{f^\prime}\left(\nabla_\mu \nabla_\nu-g_{\mu\nu}\Box\right)f^\prime+\f{C_\mu \partial_\nu \phi}{f^\prime\sqrt{-g}}.
\end{align}
One should mention that $f^\prime$ plays the role of modified gravitational constant which should be positive $f^\prime>0$. In the following, we will impose this constraint for the analysis of energy condition.

From equation \eqref{Ein} one can define the effective energy-momentum tensor as
\begin{align}\label{Teff}
T^{eff}_{\mu\nu}=\f{T_{\mu\nu}}{f^\prime}+\left(\f{f}{f^\prime}-R\right)g_{\mu\nu}+\f{2}{f^\prime}\left(\nabla_\mu \nabla_\nu-g_{\mu\nu}\Box\right)f^\prime+2\f{C_\mu \partial_\nu \phi}{f^\prime\sqrt{-g}}.
\end{align}
The Einstein's field equation is then given by
\begin{align}\label{effeq}
G_{\mu\nu}=T^{eff}_{\mu\nu}.
\end{align}
Using the Bianchi identity, one can see that the effective energy-momentum tensor is conserved, $\nabla^\mu T^{eff}_{\mu\nu}=0$.

It is worth mentioning that the same procedure of \cite{Golovnev} can be done in the present case to show that \eqref{miFRaction} is equivalent to the action
\begin{align}\label{ac2}
S=\int d^4 x\sqrt{-g}\big(f(R)+\lambda(g^{\mu\nu}\phi_\mu\phi_\nu+1)+\mathcal{L}_m\big).
\end{align}
In this form, the scalar field $\phi$ is minimally coupled to the metric and the constraint equation \eqref{cons} is added to action through the Lagrange multiplier. To obtain the field equations one should vary the above action with respect to $\lambda$, the metric tensor $g_{\mu\nu}$ and $\phi$. The results are exactly the same as \eqref{cons}, \eqref{eqg} and \eqref{eqphi}.

\section{Brans-Dicke equivalence of mimetic-$f(R)$ gravity}\label{secBD}
As is well-known, the standard $f(R)$ gravity is equivalent to the Brans-Dicke theory with the parameter $\omega=0$. Let us consider the equivalence between mimetic-$f(R)$ gravity and Brans-Dicke theory in the absence of the matter Lagrangian. The Legendre transformation for the Lagrangian \eqref{miFRaction} can be written as
\begin{align}
S=\int d^4 x \sqrt{-g(\hat{g},\phi)}\big(\Psi R(\hat{g},\phi)-V(\Psi)\big),
\end{align}
where
$$\Psi=f^\prime (R(\hat{g},\phi)), \qquad V(\Psi)=\Psi R(\hat{g},\phi)- f(R(\hat{g},\phi)),$$
where the \textit{prime} denotes derivation with respect to the argument. The above action shows that the action \eqref{miFRaction} is also equivalent to the Brans-Dicke theory with parameter $\omega=0$. The only difference between the usual Brans-Dicke theory and the above action is that the physical metric is defined by the relation \eqref{moa1}. So, we can call this theory as mimetic-Brans-Dicke theory. To obtain the field equations one should vary the action with respect to the $\Psi$, $\phi$ and $\hat{g}_{\mu\nu}$. The results are
\begin{subequations}\label{bd1}
\begin{align}
&V^\prime(\Psi)=R(g),\\
\Psi G_{\mu\nu}+(g_{\mu\nu}\Box-\nabla_\mu \nabla_\nu)\Psi+&\f{1}{2}g_{\mu\nu}V(\Psi)-\phi_\mu \phi_\nu \big(\Psi G+3\Box \phi +2 V(\Psi)\big)=0,\\
\nabla^\alpha &\big(\phi_\alpha \left(\Psi G+3\Box \phi +2 V(\Psi)\right)\big)=0,
\end{align}
\end{subequations}
together with equation \eqref{moa1}. We note that $G$ is the trace of Einstein tensor. In the above equations the covariant derivatives and also the Einstein tensor are constructed from the physical metric $g_{\mu\nu}$.  

If we use the action \eqref{ac2} for the mimetic-$f(R)$ gravity instead of \eqref{miFRaction} and performed the Legendre transformation, we would obtain
\begin{align}
S=\int d^4 x \sqrt{-g} \big[(\Psi R- V(\Psi))+\lambda(g^{\mu\nu}\phi_\mu \phi_\nu+1)\big],
\end{align}
where
$$\Psi=f^\prime(R),\qquad V(\Psi)=\Psi R -f(R).$$
It can be easily seen that the above action is the Brans-Dicke theory with $\omega=0$ which minimally couples to a constrained scalar field through the Lagrange multiplier. By varying the action with respect to the $\Psi$, $g_{\mu\nu}$ and $\phi$, one will obtain the same equations \eqref{bd1}. The $\lambda$ field equation then gives \eqref{cons}.

\section{Review on Dolgov-Kawasaki instability in mimetic-$f(R)$ theory}\label{secDK}
In this section we will obtain the Dolgov-Kawasaki criterion for mimetic-$f(R)$ gravity theory. We will use the standard procedure of \cite{insfR,noji1}. Consider the dynamics of a normal mass celestial body like the earth or sun. In this case the gravity produced by these objects is small and we can assume that locally the metric is very closed to the Minkowski one. However, the curvature can be non-zero. Suppose that the Ricci scalar differs from its background value $R_0$ (produced by the celestial body) by $R=R_0+R_1$, where $R_1$ is a perturbation around $R_0$. Also assume that the scalar field is perturbed from its background value $\phi_0(t)=t$ as $\phi(t,\vec{x})=\phi_0(t)+\phi_1(t,\vec{x})$. In order to be consistent with solar system tests, we also assume that the function $f(R)$ is very close to the Ricci scalar
$$f(R)=R+\epsilon\psi(R).$$
With these assumptions, one can expand the various derivatives of the function $f(R)$ as
\begin{align}
f(R)&=R_0+R_1+\epsilon\psi_0+\epsilon\psi_0^\prime R_1+...,\\
f^\prime(R)&=1+\epsilon\psi_0^\prime+\epsilon\psi_0^{\prime\prime}R_1+...,\\
f^{\prime\prime}(R)&=\epsilon\psi_0^{\prime\prime}+\epsilon\psi_0^{\prime\prime\prime}R_1+...,\\
f^{\prime\prime\prime}(R)&=\epsilon\psi_0^{\prime\prime\prime}+\epsilon\psi_0^{\prime\prime\prime\prime}R_1+...,
\end{align}
where $\psi_0\equiv\psi(R_0)$. Because the metric is very close to the Minkowski metric, one can write
$$\Box R=-\partial_t^2 R+\vec{\nabla}^2 R,\qquad \nabla_\mu R\nabla^\mu R=-(\partial_t R)^2+(\partial_i R)^2.$$
One should note that we have two dynamical and one constraint equations. In order to study the perturbations of the Ricci scalar we have to take the trace of metric field equation. The trace of equation \eqref{eqg} vanishes after using the constraint equation \eqref{cons}. Expanding the constraint equation up to first order gives $\partial_t\phi_1=0$ with implies that $\phi_1=\phi(\vec{x})$. So we need to consider only the scalar field equation.

Multiplying the scalar field equation \eqref{phi2} by $\partial^\mu\phi$ and using the constraint equation \eqref{cons}, one can obtain
\begin{align}
2f-Rf^\prime-3\Box f^\prime +\f12 T=-\f{C_\mu\partial^\mu\phi}{\sqrt{-g}}.
\end{align}
Expanding the above equation up to first order gives
\begin{align}
\ddot{R}_1+2\partial_t R_0\f{\psi_0^{\prime\prime\prime}}{\psi_0^{\prime\prime}}\dot{R}_1
-2\f{\psi_0^{\prime\prime\prime}}{\psi_0^{\prime\prime}}\vec{\nabla}R_0.\vec{\nabla}R_1-\vec{\nabla}^2 R_1+m_{eff}^2R_1=
V_{eff},
\end{align}
where we have defined
\begin{align}
m_{eff}^2&=\f{1}{3\epsilon\psi_0^{\prime\prime}}+\f{\psi_0^\prime}{3\psi_0^{\prime\prime}}-\f13 R_0-\f{\psi_0^{\prime\prime\prime}}{\psi_0^{\prime\prime}}(\Box R_0-\nabla_\mu R_0\nabla^\mu R_0),\\
V_{eff}&=\Box R_0-\f{C_\mu\partial^\mu\phi_1}{3\epsilon\psi_0^{\prime\prime}}-\f23\f{\psi_0}{\psi_0^{\prime\prime}}+\f13R_0\f{\psi_0^\prime}{\psi_0^{\prime\prime}}+\f{\psi_0^{\prime\prime\prime}}{\psi_0^{\prime\prime}}\nabla_\mu R_0\nabla^\mu R_0.
\end{align}
One can see that the effective potential of the mimetic theory is changed due to the presence of a constrained scalar field. 
The effective mass is dominated by its first term which implies a constraint $\psi_0^{\prime\prime}>0$ for a theory to avoid tachyonic instability. We note that $\psi_0^{\prime\prime}=f^{\prime\prime}(R_0)$. So, the present theory has the criteria
$$f^{\prime\prime}(R_0)>0,$$
to avoid Dolgov-Kawsaki instability (see also \cite{fRMDM}). We note that the same condition is hold for $f(R,T)$ theory. However, non-minimal couplings between the Ricci tensor and the energy-momentum tensor can modify the above condition \cite{fRT}. In the following we will impose this constraint for analyzing the energy conditions of the theory.
\section{Energy conditions}\label{secEC}
Let us consider the flat FRW universe with scale factor $a(t)$
\be
ds^2=-dt^2+a(t)^2\left(dx^2+dy^2+dz^2\right),
\ee
which is filled with a perfect fluid with energy-momentum tensor
\be\label{perf}
T_{\mu\nu}=(p+\rho)u_\mu u_\nu +pg_{\mu\nu},
\ee 
where $\rho$ is the energy density, $p$ is the thermodynamics pressure and $u_\mu$‌ is the fluid 4-velocity field. In this case, from equation \eqref{cons}, one can obtain
\be
\phi(t)=t+c
\ee
where $c$ is an integration constant. The scalar field $\phi$ always appear with derivative, hence the theory has a shift symmetry on $\phi$, so we can set $c=0$ without loss of generality. Equation \eqref{phi2} implies that only the temporal component of $C_\mu$ be non-zero and given by
\begin{align}\label{C0}
C_{0}=\f{a^3}{2}\big(3p-\rho+4f-12f^\prime (\dot{H}+2H^2)+18H\dot{f}^\prime+6\ddot{f}^\prime\big)=const.
\end{align} 
Here, $H=\dot{a}/a$ is the Hubble parameter and \textit{dot} represents time derivative. The effective energy-density and pressure of the mimetic-$f(R)$ gravity theory can be obtained from equation \eqref{Teff} as
\begin{align}\label{rop}
\rho_{eff}&=12H^2+6\dot{H}+\f{1}{f^\prime}\bigg(\rho-f-6H\dot{f}^\prime+\f{2C_0}{a^3}\bigg),\nonumber\\
p_{eff}&=-12H^2-6\dot{H}+\f{1}{f^\prime}\bigg(p+f+4H\dot{f}^\prime+2\ddot{f^\prime}\bigg),
\end{align}
\subsection{The Raychaudhuri Equation}
The energy conditions play an important role in understanding the gravitational nature of the gravitating system. In general relativity, we have four explicit forms of energy conditions, namely: Strong (SEC), null (NEC), dominant (DEC) and weak energy condition (WEC) \cite{haw}.
Among these four energy conditions, the SEC and NEC can be obtained using the well-known Raychaudhuri equation, from the fact that the gravitational force is attractive. In the case of a time-like and null geodesic congruence with tangent vector fields $u^\mu$ and $k^\mu$ respectively, the Raychaudhuri equation can be written as
\begin{align}\label{ec}
\f{d\theta}{d\tau}&=-\f13\theta^2-\sigma_{\mu\nu}\sigma^{\mu\nu}+\omega_{\mu\nu}\omega^{\mu\nu}-R_{\mu\nu}u^\mu u^\nu,\\
\f{d\theta}{d\lambda}&=-\f12\theta^2-\sigma_{\mu\nu}\sigma^{\mu\nu}+\omega_{\mu\nu}\omega^{\mu\nu}-R_{\mu\nu}k^\mu k^\nu,
\end{align}
where $\theta$ is the expansion scalar, $\omega_{\mu\nu}$ is the rotation tensor, $\sigma_{\mu\nu}$ is the shear tensor and $\tau$ and $\lambda$‌ are the affine parameters of time-like and null geodesics respectively. In the above expression we have assumed that all non-gravitational forces are absent \cite{21maghale}.

In order to obtain the energy conditions, let us assume that the quadratic terms in the expressions \eqref{ec} vanish. In this case the congruence is rotation free and have infinitesimal distortions. One can then integrate \eqref{ec} to obtain $\theta=-\tau R_{\mu\nu}u^\mu u^\nu$ and $\theta=-\lambda R_{\mu\nu}k^\mu k^\nu$, for time-like and null congruences respectively. Noting that the gravitational force is attractive, i.e. $\theta<0$, one obtains
\begin{align}
R_{\mu\nu}u^\mu u^\nu\geq0,\qquad R_{\mu\nu}k^\mu k^\nu\geq0.
\end{align}
In Einstein's general relativity, one can use the metric field equations $G_{\mu\nu}=T_{\mu\nu}$ to obtain
\begin{align}
\left(T_\mu\nu-\f12g_{\mu\nu}T\right)u^\mu u^\nu\geq0,\qquad \left(T_\mu\nu-\f12g_{\mu\nu}T\right)k^\mu k^\nu\geq0.
\end{align}
In the case of perfect fluid \eqref{perf}, the energy conditions reduce to the Strong and Null energy conditions
\begin{align}
\textmd{SEC:}&\quad \rho+3p\geq0 \quad\&\quad \rho+p\geq0,\nonumber\\
\textmd{NEC:}&\quad \rho+p\geq0.
\end{align}
Besides these energy conditions, we have two other energy conditions known as the weak energy condition (WEC) and the Dominant energy condition (DEC). The WEC implies that the local energy density as measured by any time-like observer be positive. One can Write the WEC as
\begin{align}
\textmd{WEC:}\quad T_{\mu\nu}u^\mu u^\nu\geq0.
\end{align}
Also, the DEC states that the locally measured energy density is positive and the energy flux is time-like or null. One can write the DEC as
\begin{align}
\textmd{DEC:}\quad T_{\mu\nu}u^\mu u^\nu\geq0\quad \& \quad T_{\mu\nu}u^\mu \quad \textmd{is timelike or null}.
\end{align}
assuming a perfect fluid, one can obtain
\begin{align}
\textmd{WEC:}&\quad \rho\geq0 \quad\&\quad \rho+p\geq0,\nonumber\\
\textmd{DEC:}&\quad \rho\geq0\quad \rho\pm p\geq0.
\end{align}
As we have mentioned earlier, in this paper we want to obtain conditions in which the universe undergoes an accelerated expansion. In this case the SEC should be violated but other energy conditions should remain satisfied. Considering the above conditions, one can deduce that the condition $\rho+3p\geq0$ should be violated. One can summarize all the conditions above as a list 
\begin{flalign} \label{list}
	&\qquad\bullet\textmd{(I)}:\quad~~ \rho\geq 0,&\nonumber\\ 
	&\qquad\bullet\textmd{(II)}:\quad~ \rho+p\geq 0,&\nonumber\\
	&\qquad\bullet\textmd{(III)}:\quad \rho+3p\geq 0,&\nonumber\\
	&\qquad\bullet\textmd{(IV)}:\quad \rho-p\geq 0.&
\end{flalign}
The above arguments implies that the condition (III) should be violated while the other conditions should remain satisfied.
\subsection{Energy conditions in mimetic-$f(R)$ gravity}
In this section we want to write the energy conditions in the mimetic-$f(R)$ gravity theory. In equation \eqref{effeq}, the metric field equations of the mimetic-$f(R)$ gravity is formulated in a way that is similar to Einstein's field equation. The only difference is that the energy-momentum tensor is replaced by the effective energy-momentum tensor which is now contain the effects of both ordinary matter and geometry. One can then think that the energy conditions can be written in a similar way as Einstein's theory, by replacing the energy density and pressure with their effective quantities. In this sense, one can write the conditions \eqref{list} in mimetic-$f(R)$ theory as
\begin{flalign} \label{listmim}
&\qquad\bullet\textmd{(I)}:~~\quad \rho_{eff}\geq 0,&\nonumber\\ 
&\qquad\bullet\textmd{(II)}:~\quad \rho_{eff}+p_{eff}\geq 0,&\nonumber\\
&\qquad\bullet\textmd{(III)}:\quad \rho_{eff}+3p_{eff}\geq 0,&\nonumber\\
&\qquad\bullet\textmd{(IV)}:\quad \rho_{eff}-p_{eff}\geq 0.&
\end{flalign}
Using equation \eqref{rop}, one can write the conditions \eqref{listmim} as
\begin{flalign} 
&\qquad\bullet\textmd{(I)}:\qquad 12H^2+6\dot{H}+\f{1}{f^\prime}\bigg(\rho-f-6H\dot{f}^\prime+\f{2C_0}{a^3}\bigg)\geq 0,&\nonumber\\ &\qquad\bullet\textmd{(II)}:\qquad-24H^2-12\dot{H}+\f{1}{f^\prime}\bigg(\rho+3p+2f+6H\dot{f}^\prime+6\ddot{f}^\prime+\f{2C_0}{a^3}\bigg)> 0,&\nonumber\\
&\qquad\bullet\textmd{(III)}:\qquad\f{1}{f^\prime}\bigg(\rho+ p+2\ddot{f}^\prime-2H\dot{f}^\prime+\f{2C_0}{a^3}\bigg)\geq 0,&\nonumber\\
&\qquad\bullet\textmd{(IV)}:\qquad 24H^2+12\dot{H}+\f{1}{f^\prime}\bigg(\rho-p-2f-2\ddot{f}^\prime-10H\dot{f}^\prime+\f{2C_0}{a^3}\bigg)\geq 0.&
\end{flalign}
The above conditions ensure that the gravitational force remains attractive and the energy density of the universe remain positive. However, as we have mentioned earlier,  if one wants to consider special cases where the universe encounter an accelerated expansion, one should search for conditions where (III) is violated. In the following we will concentrate on the conditions where all but (III) are satisfied. In the following we will call this special case ``the accelerated expanding condition". Note that in the above calculations we check that the conditions $f^\prime>0$ and $f^{\prime\prime}>0$ are satisfied ensuring that the gravitational constant is positive and the Dolgov-Kawasaki instability is avoided as was discussed in the previous sections.

In order to put a better constraint on the parameters of the theory, it is helpful to write the energy conditions in terms of the Hubble parameter, $H$, deceleration, jerk and snap parameters with definitions
\begin{align}
q=-\f{1}{H^2}\f{\ddot{a}}{a},\quad j=\f{1}{H^3}\f{\dddot{a}}{a},\quad s=\f{1}{H^4}\f{\ddddot{a}}{a}.
\end{align}
The present day values of the deceleration parameter,the jerk and the Hubble parameter are $q_0=-0.30\pm0.16$ with the best fit $q_0=-0.24$, $j_0=-4.62\pm 1.74$ with the best fit $-4.82$ and $H_0=71.16\pm3.08$ with the best fit $H_0=71.65$  \cite{plank}. From the constraint equation \eqref{C0}, one can obtain the snap parameter in terms of $q_0$, $H_0$ and $j_0$ and $C_0$.
The constraint equation \eqref{C0} can be written in terms of cosmological parameters as
\begin{align}\label{eq1}
C_0=54\bigg(\f{1}{27}f-\f{1}{108}(\rho_0-3p_0)+2H_0^6(j_0-q_0-2)^2f^{\prime\prime\prime}+\f13 H_0^4(q_0^2+3j_0+5q_0+s_0)f^{\prime\prime}+\f19 H_0^2(q_0-1)f^\prime\bigg),
\end{align}
where we have set $a_0=1$.
condition (I) is independent of the snap parameter and can be written as
\begin{align}
\f{1}{f^\prime}\bigg(\rho_0-f+6H_0^2(1-q_0)f^\prime-36H_0^4(j_0-q_0-2)f^{\prime\prime}+2C_0\bigg)\geq 0.
\end{align}
In the other three energy conditions the snap parameter appears, which can be substituted with other parameters using equation \eqref{eq1}, with the result
\begin{flalign} 
&\qquad\bullet\textmd{(II)}:\qquad\f{1}{3f^\prime}\bigg(4\rho_0-4f-12H_0^2(q_0-1)f^\prime-144H_0^4(j_0-q_0-2)f^{\prime\prime}+6C_0\bigg)\geq0,&\nonumber\\
&\qquad\bullet\textmd{(III)}:\qquad\f{1}{f^\prime}\bigg(2\rho_0-2f-72H_0^4(j_0-q_0-2)f^{\prime\prime}+2C_0\bigg)<0,&\nonumber\\
&\qquad\bullet\textmd{(IV)}:\qquad \f{1}{3f^\prime}\bigg(2\rho_0-2f-24H_0^2(q_0-1)f^\prime-72H_0^4(j_0-q_0-2)f^{\prime\prime}+6C_0\bigg)\geq0.&
\end{flalign}
where we have changed the inequality of condition (III), signaling its violation.
Note that by setting $16\pi G=1=c$ in this paper, the Hubble parameter has dimension $L^{-1}$, where $L$ represents the dimension of length. Also the energy density $\rho$, the constant $C_0$ and the function $f(R)$ have dimensions $L^{-2}$. One can then write the above conditions as
\begin{flalign}\label{enco}
&\qquad\bullet\textmd{(I)}:\qquad~~ \f{1}{g^\prime}\bigg(6\Omega_m-g+6(1-q_0)g^\prime-36(j_0-q_0-2)g^{\prime\prime}+2c_0\bigg)\geq 0,&\nonumber\\ 
&\qquad\bullet\textmd{(II)}:\qquad\f{1}{3g^\prime}\bigg(24\Omega_m-4g-12(q_0-1)g^\prime-144(j_0-q_0-2)g^{\prime\prime}+6c_0\bigg)\geq0,&\nonumber\\
&\qquad\bullet\textmd{(III)}:\qquad\f{1}{g^\prime}\bigg(12\Omega_m-2g-72(j_0-q_0-2)g^{\prime\prime}+2c_0\bigg)<0,&\nonumber\\
&\qquad\bullet\textmd{(IV)}:\qquad \f{1}{3g^\prime}\bigg(12\Omega_m-2g-24(q_0-1)g^\prime-72(j_0-q_0-2)g^{\prime\prime}+6c_0\bigg)\geq0.&
\end{flalign}
where we have defined dimensionless quantities
\begin{align}
c_0\equiv \f{C_0}{H_0^2},\qquad \Omega_m\equiv\f{\rho_0}{\rho_c}=\f{\rho}{6H_0^2},\qquad r\equiv \f{R}{H_0^2},\qquad g\equiv \f{f}{H_0^2}.
\end{align}
We should note that $\Omega_m$‌ is the current density of baryonic plus dark matter. This is because we have assumed that the mimetic-$f(R)$ theory can explain the dark energy content of the universe, and so from equation \eqref{Teff}, $T_{\mu\nu}$ corresponds to the baryonic plus dark matter content of the universe. In the above relations \textit{prime} represents derivative with respect to $r$. 

In the following we will use the best fit values for the parameters $q_0$‌ and $j_0$ and set $\Omega_{m0}=0.27$ which is the density of baryonic plus dark matter of the universe. This value is in accordance with the values of $H_0$, $q_0$ and $j_0$ we have used here (see \cite{plank} for the analysis we have used in this paper for the cosmological parameters).
\section{Some special cases}\label{some}
In this section we will consider some special cases for $f(R)$ in order to investigate the energy conditions quantitatively.
\subsection{The Case $f=R+\alpha R^n$}\label{sub1}
Note that the constant $\alpha$ has dimension $L^{2(1-n)}$. Introducing $\beta=H_0^{2(n-1)}\alpha$, one obtains
$g(r)=r+\beta r^n.$
The conditions \eqref{enco} can be written as
\begin{flalign}\label{powe}
&\qquad\bullet\textmd{(I)}:\quad~~~~ \beta (4.27n^2-3.27n-1.00)(7.44)^n+2c_0+1.62\geq 0,&\nonumber\\ 
&\qquad\bullet\textmd{(II)}:\quad~~~ \beta (5.71n^2-5.04n-1.33)(7.44)^n+2c_0-2.80\geq0,&\nonumber\\
&\qquad\bullet\textmd{(III)}:\qquad\beta (8.56n^2-8.56n-2.00)(7.44)^n+2c_0-11.64<0,&\nonumber\\
&\qquad\bullet\textmd{(IV)}:\qquad\beta (2.80n^2-1.52n+0.67)(7.44)^n+2c_0+6.04\geq0.&
\end{flalign}
and the conditions for positivity of the first and second derivatives of $f(R)$ can be summarized as
$$1+0.13 \beta n(7.44)^n>0 \qquad \&\qquad \beta n(n-1)(7.44)^n>0.$$

\begin{figure}[h!]
\centering
\includegraphics[scale=0.37]{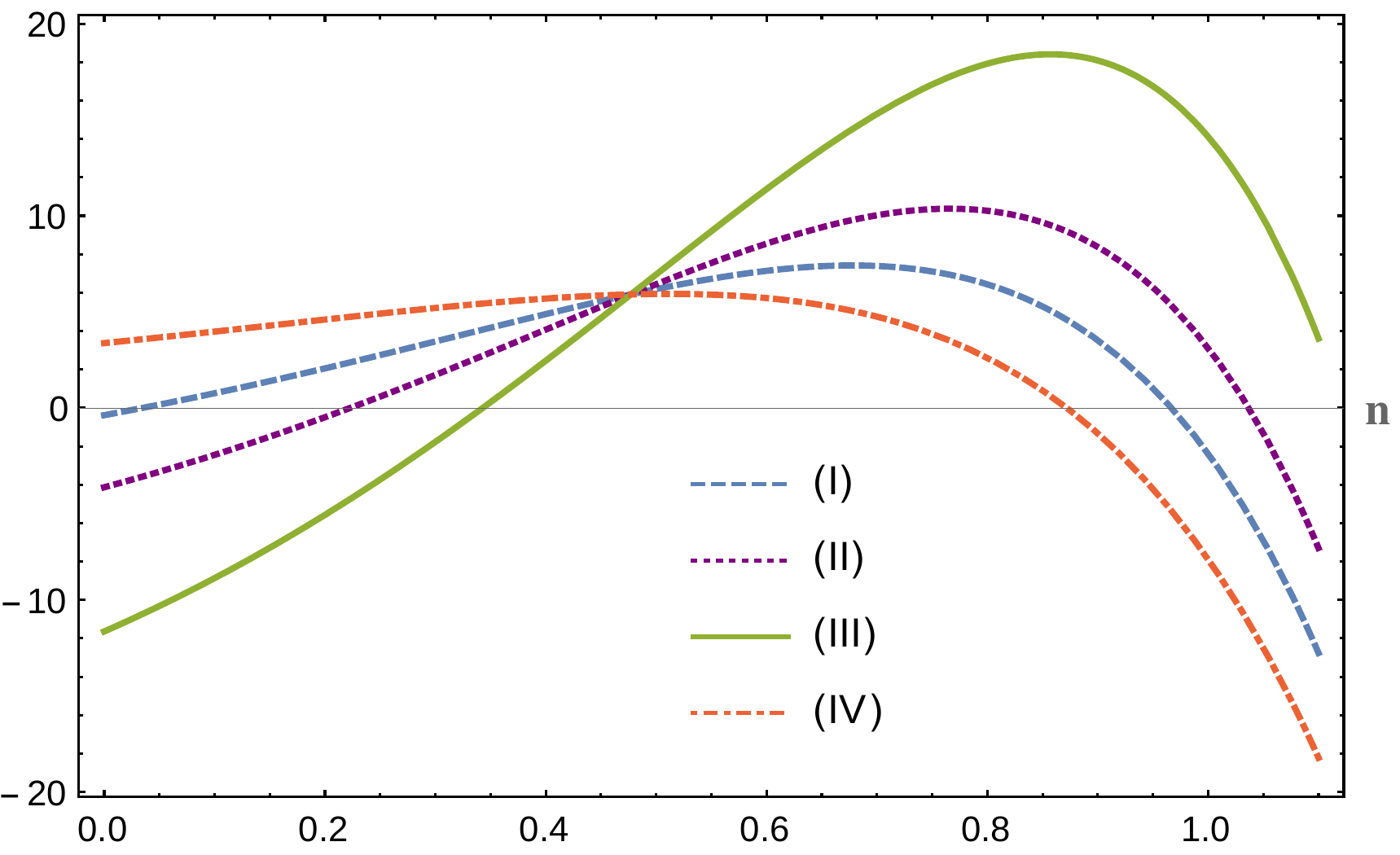}
\caption{\label{fig1} The conditions \eqref{powe} for $c_0=-2$ and $\beta=-2$ in the case $g(r)=r+\beta r^n$. As one can see, the only possible values for n lie between the point where (II) vanishes and the point where (III) vanishes.}
\end{figure} 
Let us first consider the case $\beta<0$. The positivity of $f^{\prime\prime}$ implies that $n\in(0,1)$. In figure \eqref{fig1} we have plotted the conditions \eqref{powe} for $c_0=-2$ and $\beta=-2$. Other values for $c_0$‌ and $\beta$ will also result in a qualitatively same figure. As one can see from the figure, there exists always an intersection at $n=n_{int}$ where all the conditions become equal. For $n>n_{int}$ the condition (III) becomes greater than other conditions and hence the accelerated expanding condition can not be satisfied. So we should have $n<n_{int}$. One can check that at $n_{int}$ the value of condition (III) is always positive and all the conditions are increasing functions of $n$. Also, the condition (II) is less than conditions (I) and (IV) for $n<n_{int}$. The accelerated expanding condition will then be satisfied in the range  $n\in(n_2,n_3)$, where $n_2$ and $n_3$ is a point in which the condition (II) and (III) become zero respectively.
\begin{figure}[h!]
	\centering
	\includegraphics[scale=0.4]{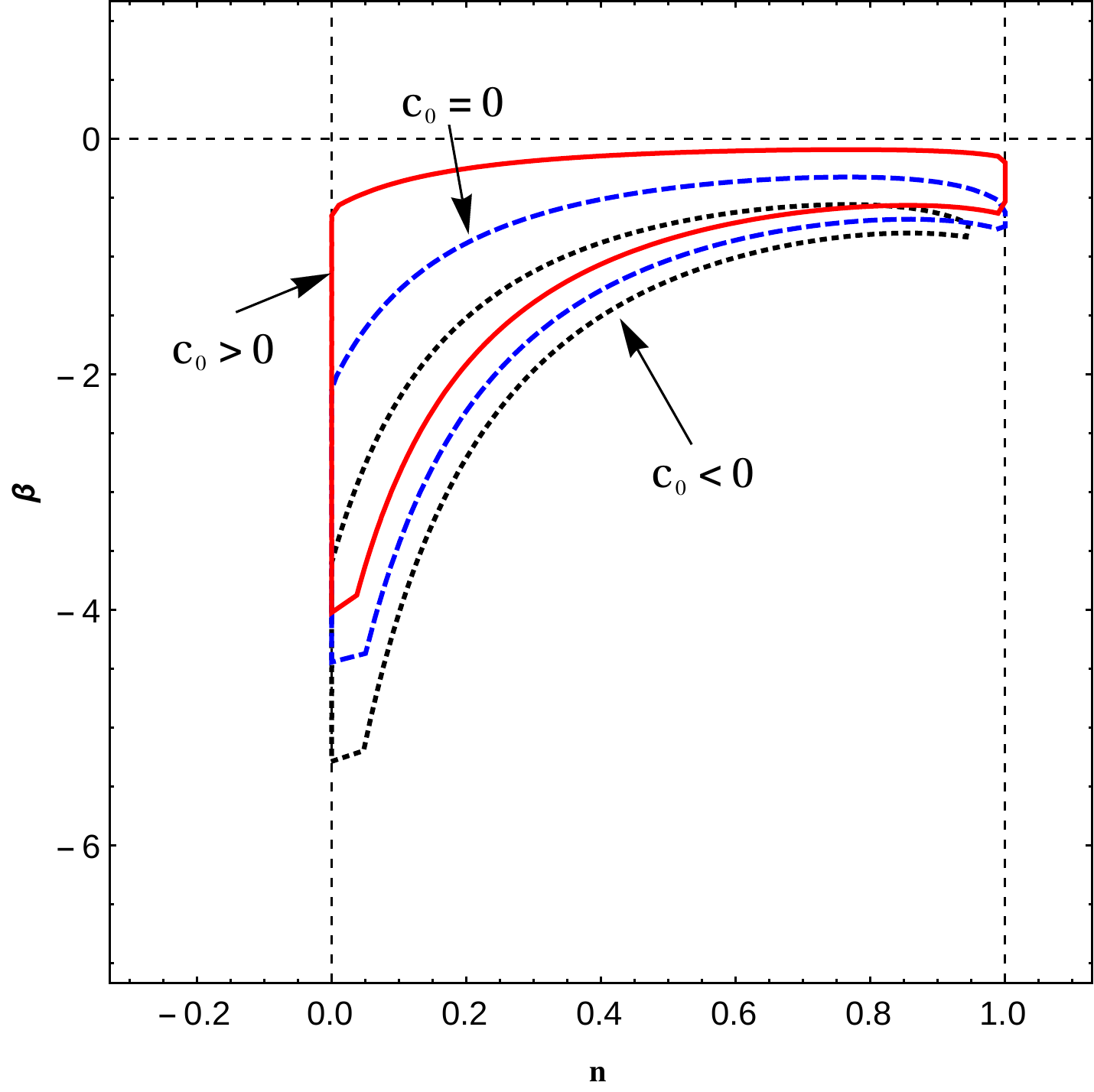}
	\caption{\label{fig2} The allowed range of parameters $n$ and $\beta>0$ in the case $c_0=3$ for $g(r)=r+\beta r^n$.}
\end{figure} 
In figure \eqref{fig2} we have plotted the allowed range of $\beta$ and $n$‌ parameters for $c_0=-1,0,+1$. One can see from the figures that for positive values of $c_0$ greater values of $\beta$‌ can be chosen. Also, for negative values of $c_0$ the allowed values of $\beta$‌ becomes smaller.

\begin{figure}[h!]
	\centering
	\includegraphics[scale=0.35]{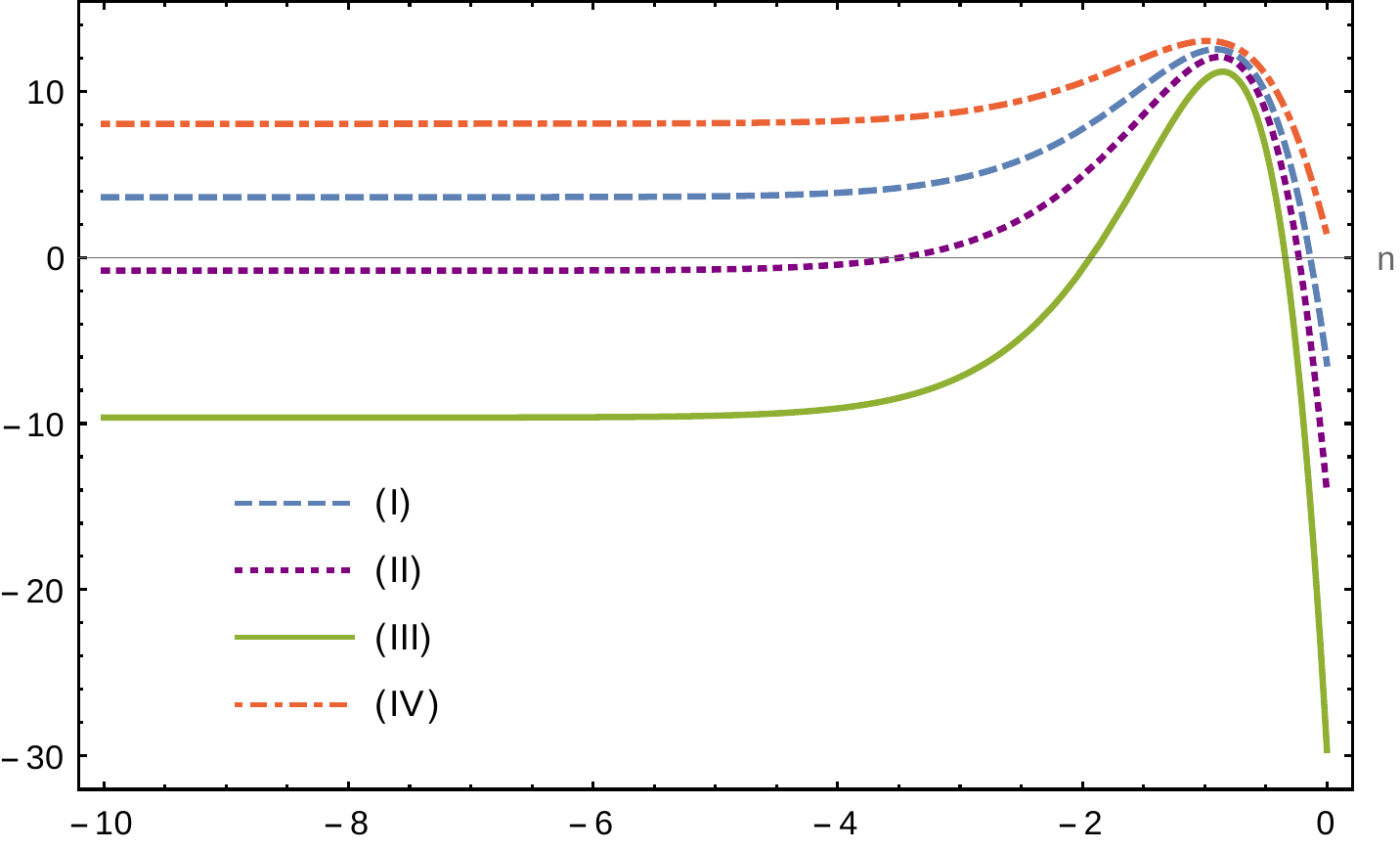}
	\includegraphics[scale=0.35]{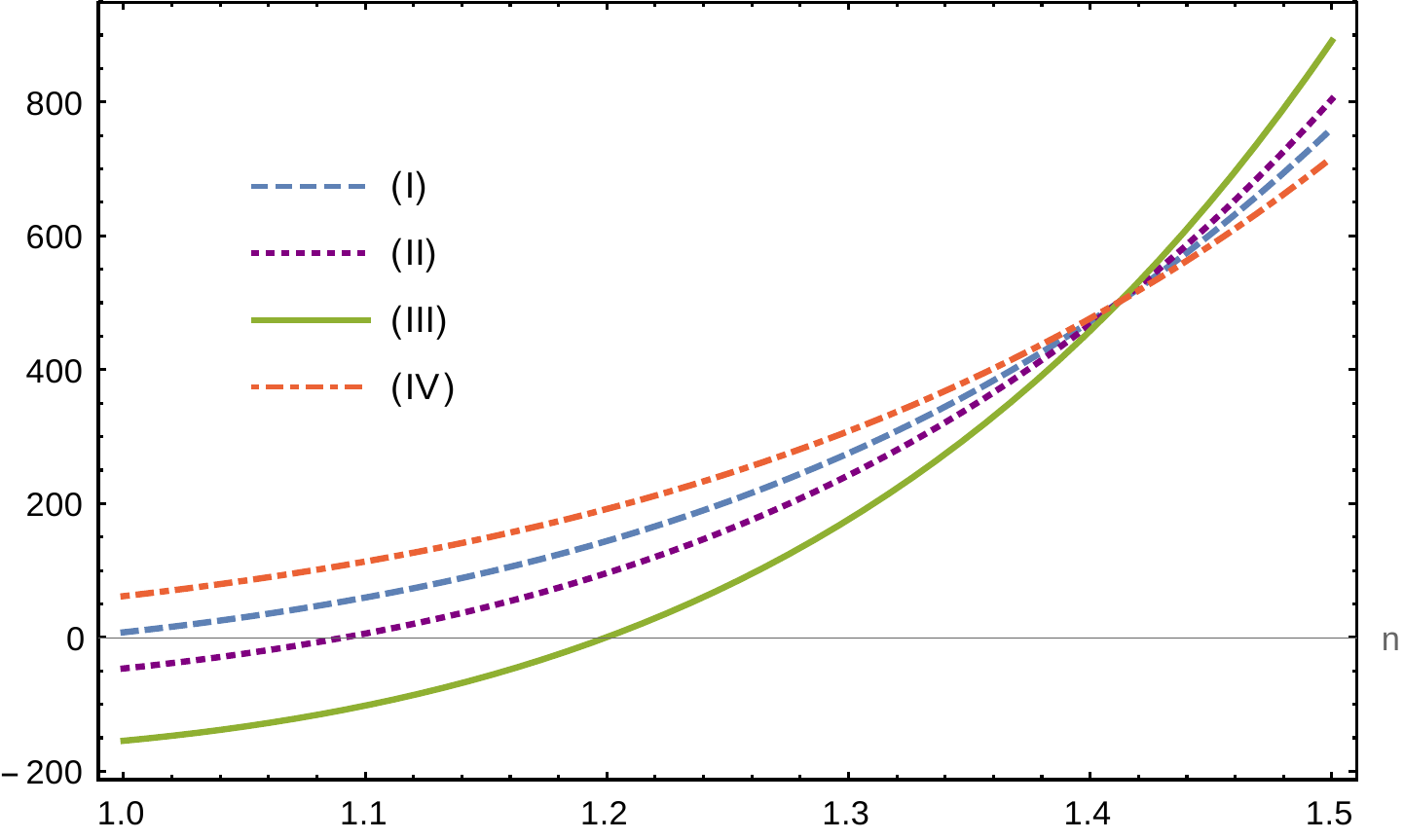}
	\caption{\label{fig3} The conditions \eqref{powe}‌for $c_0=3$ and $\beta=10$ in the case $g(r)=r+\beta r^n$. The left figure correspond to $n\in(-\infty,0)$ and the right figure corresponds to $n\in(1,+\infty)$.}
\end{figure}

In the case $\beta>0$, the allowed range of n is $n\notin(0,1)$. Let us decompose this range into negative values $n\in(-\infty,0)$, and positive values $n\in(1,+\infty)$. In figure \eqref{fig3} we have plotted all the conditions \eqref{powe} as a function of $n$‌ for $c_0=3$ and $\beta=10$.
In the range $n\in(-\infty,0)$, one can see that all the conditions approach to a constant at minus infinity and the condition (III) has the smallest value among them. One can see from the conditions \eqref{powe} that different values of $c_0$ moves all the curves in the vertical axis. So, the allowed range for $n$‌ is $n\in(n_2,n_3)$ where $n_2$‌ and $n_3$‌ are the values of $n$ where the conditions (II) and (III) becomes zero respectively. There exist a special case $c_0=1.4$ where the condition (II) approaches zero at minus infinity. In this case the condition (III) is negative in the range $n\in(-\infty,n_3)$‌ and the other conditions become positive.

In the range $n\in(1,+\infty)$, the qualitative behavior of the conditions are the same as in the $\beta<0$ case. In this case, there is an intersection point where all the conditions becomes equal. The allowed range is again occurs in $n\in(n_2,n_3)$ where $n_2$ and $n_3$ are the points where conditions (II) and (III) vanishes respectively.

\begin{figure}[h!]
	\centering
	\includegraphics[scale=0.45]{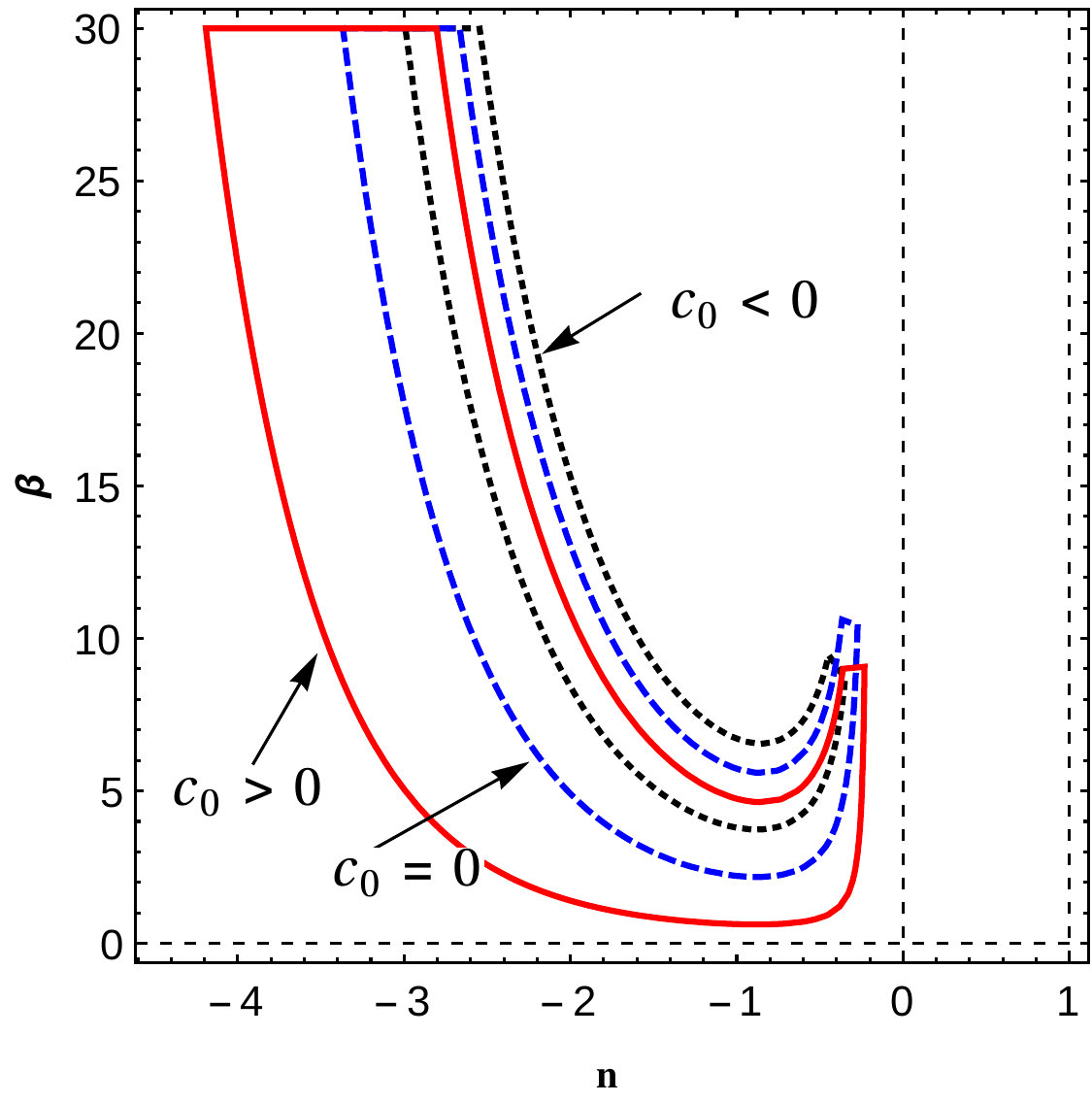}\qquad
	\includegraphics[scale=0.443]{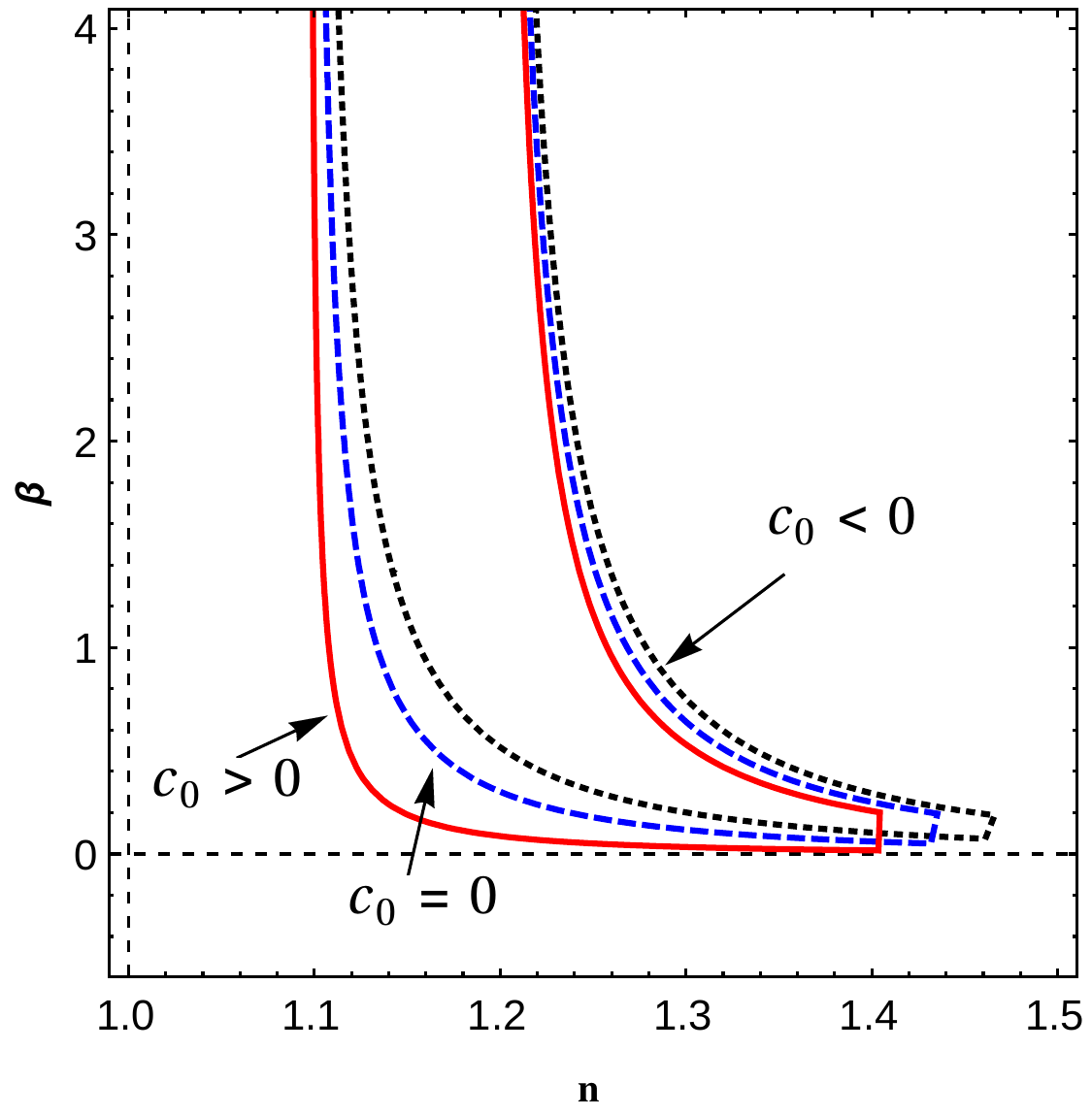}
	\caption{\label{fig4} The allowed range of parameters $n$ and $\beta$ for $c_0=-1,0,+1$ in the case $g(r)=r+\beta r^n$. The left figure correspond to $n\in(-\infty,0)$ and the right figure corresponds to $n\in(1,+\infty)$.}
\end{figure}

In figure \eqref{fig4} we have plotted the allowed parameter space for $c_0=-1,0,+1$ for both sub-ranges of $n$. One can see that for a given value of $\beta$, positive values of $c_0$ can take smaller values of $n$ and negative values of $c_0$ can take greater values of $n$.
\begin{figure}[h!]
	\centering
	\includegraphics[scale=0.4]{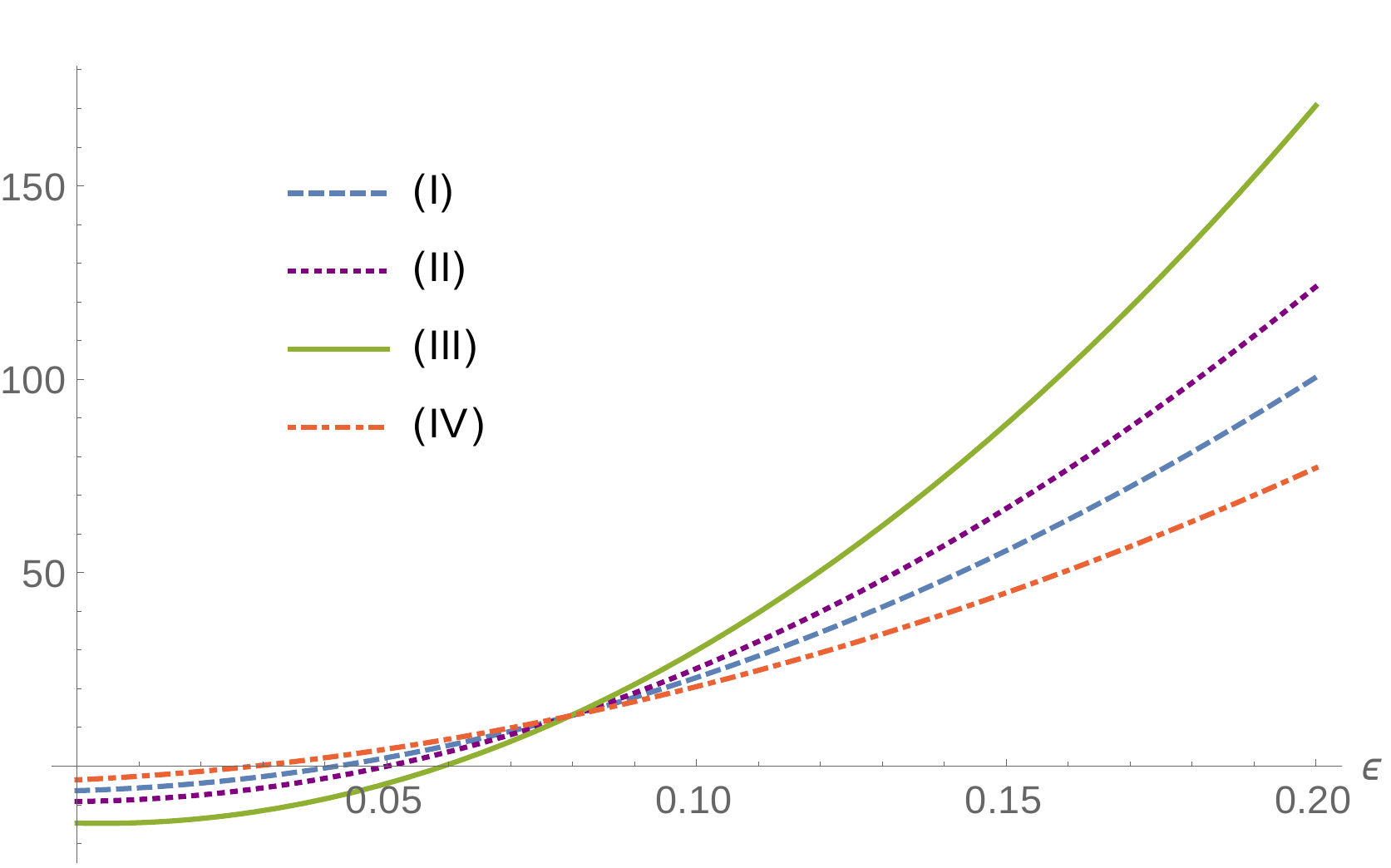}
	\caption{\label{fig5}The conditions \eqref{exp} in the case of $\eta=10$ and $c_0=1$ for $g(r)=\eta e^{\epsilon r}$. Negative values of $c_0$ will also produce qualitatively the same figure.}
\end{figure}
\subsection{The Case $f=\alpha e^{\beta R}$}
Introducing $\eta=\alpha/H_0^2$ and $\epsilon=H_0^2\beta$, one can write
$g(r)=\eta e^{\epsilon r}.$
In this case the positivity of $f^\prime$ implies that $\eta\epsilon>0$ and the positivity of $f^{\prime\prime}$ results in $\eta\epsilon^2>0$. So, both $\epsilon$ and $\eta$ parameters should be positive.
With this in hand, the conditions \eqref{enco} can be simplified to
\begin{flalign} \label{exp}
&\qquad\bullet\textmd{(I)}:\qquad~~ (236.88\epsilon^2+7.44\epsilon-1.00)\eta+(2c_0+1.62)e^{-7.44\epsilon}\geq 0,&\nonumber\\ 
&\qquad\bullet\textmd{(II)}:\qquad~(315.84\epsilon^2+4.96\epsilon-1.33)\eta+(2c_0+2.16)e^{-7.44\epsilon}\geq0,&\nonumber\\
&\qquad\bullet\textmd{(III)}:\qquad(437.76\epsilon^2-2.00)\eta+(2c_0+3.24)e^{-7.44\epsilon}<0,&\nonumber\\
&\qquad\bullet\textmd{(IV)}:\qquad (157.92\epsilon^2+9.92\epsilon-0.67)\eta+(2c_0+1.08)e^{-7.44\epsilon}\geq0.&
\end{flalign}
In figure \eqref{fig5}, we have plotted all the conditions \eqref{exp} as a function of $\epsilon$‌ for $c_0=1$ ‌and $\eta=10$. We should note that negative value of $c_0$‌ will also produce qualitatively the same figure. One can see that there is  an intersection point at $\epsilon_{int}$‌ where all the conditions become equal. For large and negative values of $c_0$ and large values of $\eta$, all the conditions are negative at the intersection point. In this case there is no allowed range in which the accelerated expanding condition is satisfied. However, for values $c_0\geq0$, the intersection point occurs when all the conditions are positive. In this case the allowed range is $\epsilon\in(\epsilon_2,\epsilon_3)$, where $\epsilon_2$ and $\epsilon_3$ are the values at which conditions (II) and (III) vanishes respectively. Also, for small values of $\eta$, there is no intersection point and the condition (III) is always greater than other conditions. In this case there is no allowed range where the accelerated expanding condition is satisfied. 
\begin{figure}[h]
\centering
\includegraphics[scale=0.4]{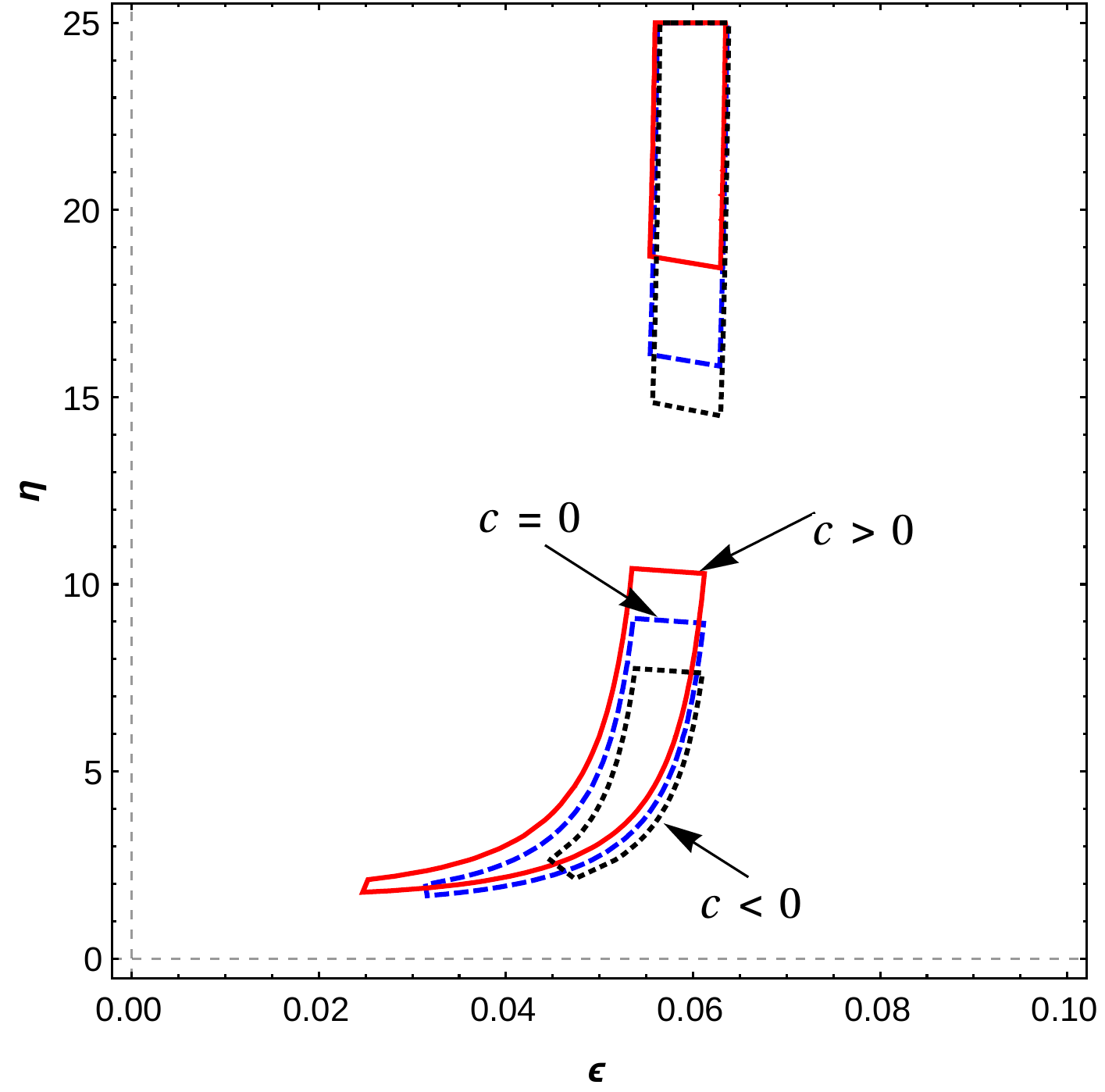}
\caption{\label{fig6} The allowed range of parameters $\epsilon$ and $\eta$ for $c_0=0.2$ (the Solid line), $c_0=0$ (the Dashed line) and $c_0=-0.2$ (the Dotted line) for $g(r)=\eta e^{\epsilon r}$.}
\end{figure}

In figure \eqref{fig6}, we have plotted the allowed parameter space of this case for $c_0=-0.2,0,+0.2$. One can see that for all choices of $c_0$, the allowed parameters contains arbitrary large values for $\eta$, but the values of $\epsilon$‌ always lies in $\epsilon\in(0.02,0.08)$. 
\subsection{The Case $f=R+\alpha R^n e^{\beta R}$}
By defining $\epsilon=H_0^\beta$ and $\eta=H_0^{2(n-1)}\alpha$ one can obtain $g(r)=r+\eta r^n e^{\epsilon r}$. The energy conditions in this case can be written as
\begin{flalign} \label{mixed}
&\qquad\bullet\textmd{(I)}:\quad~~ \eta  e^{7.44 \epsilon +2.01 n} \big(236.38 \epsilon^2+4.28 n^2+7.44\epsilon -3.28n -63.68 \epsilon n-1\big)+2c_0+1.62\geq 0,&\nonumber\\ 
&\qquad\bullet\textmd{(II)}:\qquad\eta  e^{7.44 \epsilon +2.01 n} \big(315.84 \epsilon^2+5.71 n^2+4.96\epsilon-5.04n+84.90 \epsilon n-1.33\big)+2c_0-2.80\geq0,&\nonumber\\
&\qquad\bullet\textmd{(III)}:\qquad\eta  e^{7.44 \epsilon +2.01 n} \big(473.76 \epsilon ^2+8.56n^2+127.35\epsilon  n-8.56 n-2\big)+2c_0-11.64<0,&\nonumber\\
&\qquad\bullet\textmd{(IV)}:\qquad \eta  e^{7.44 \epsilon +2.01 n} \big( 157.92\epsilon^2+2.85 n^2+9.92\epsilon-1.52n +42.45\epsilon n-0.67\big)+2c_0+6.04\geq0.&
\end{flalign}
Also the positivity of $f^\prime$ and $f^{\prime\prime}$ is reduced to
\begin{align}
&\eta  e^{7.44 \epsilon +2.01n} (\epsilon +0.13 n)+1>0,\\
& \eta  \left(\epsilon ^2- n^2+0.27 \epsilon  n- n\right)>0.
\end{align}
where we have used $r^n=e^{\ln r}$ for simplifying the equations.
\begin{figure}
\centering
\includegraphics[scale=0.45]{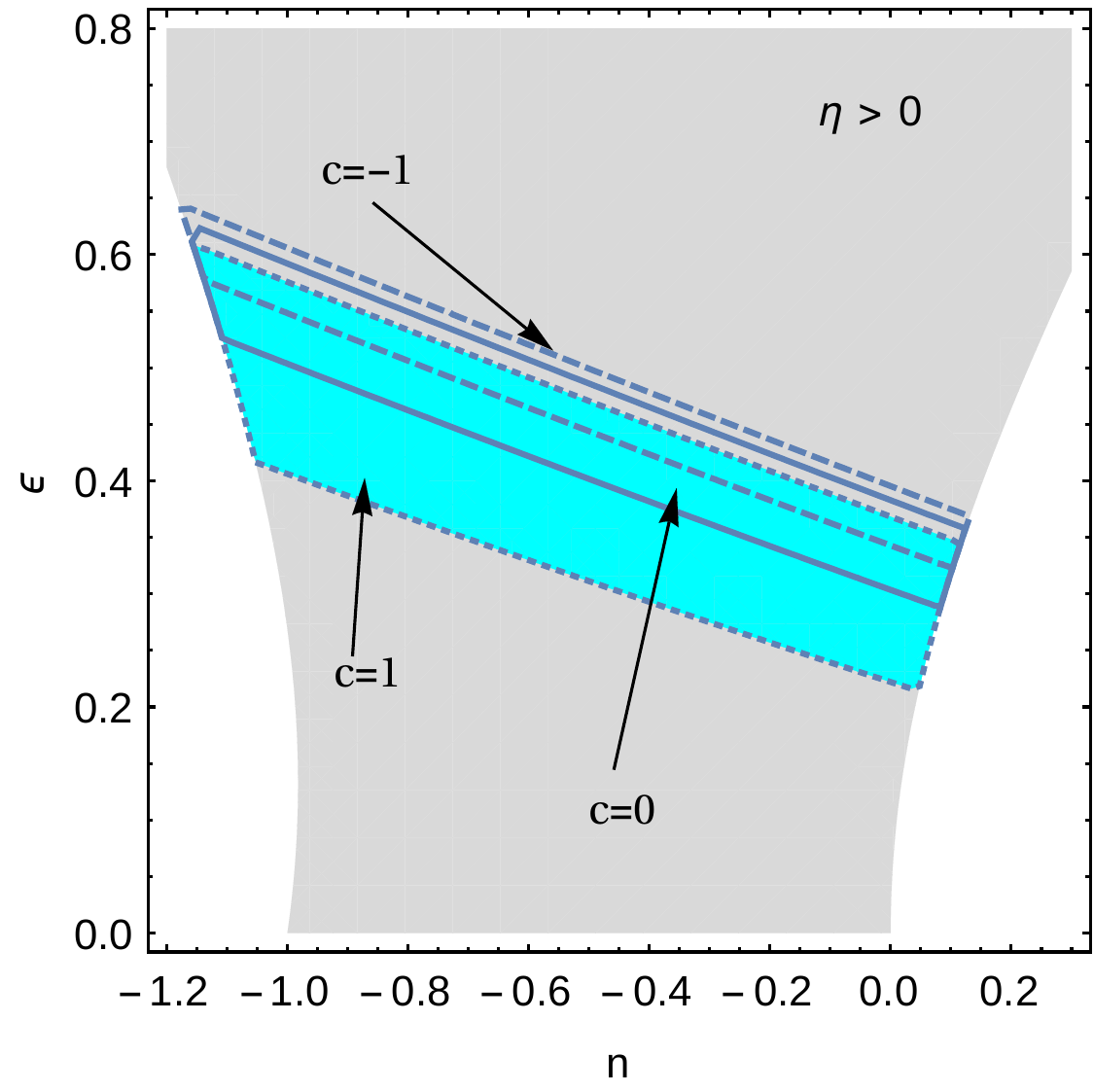}\quad
\includegraphics[scale=0.463]{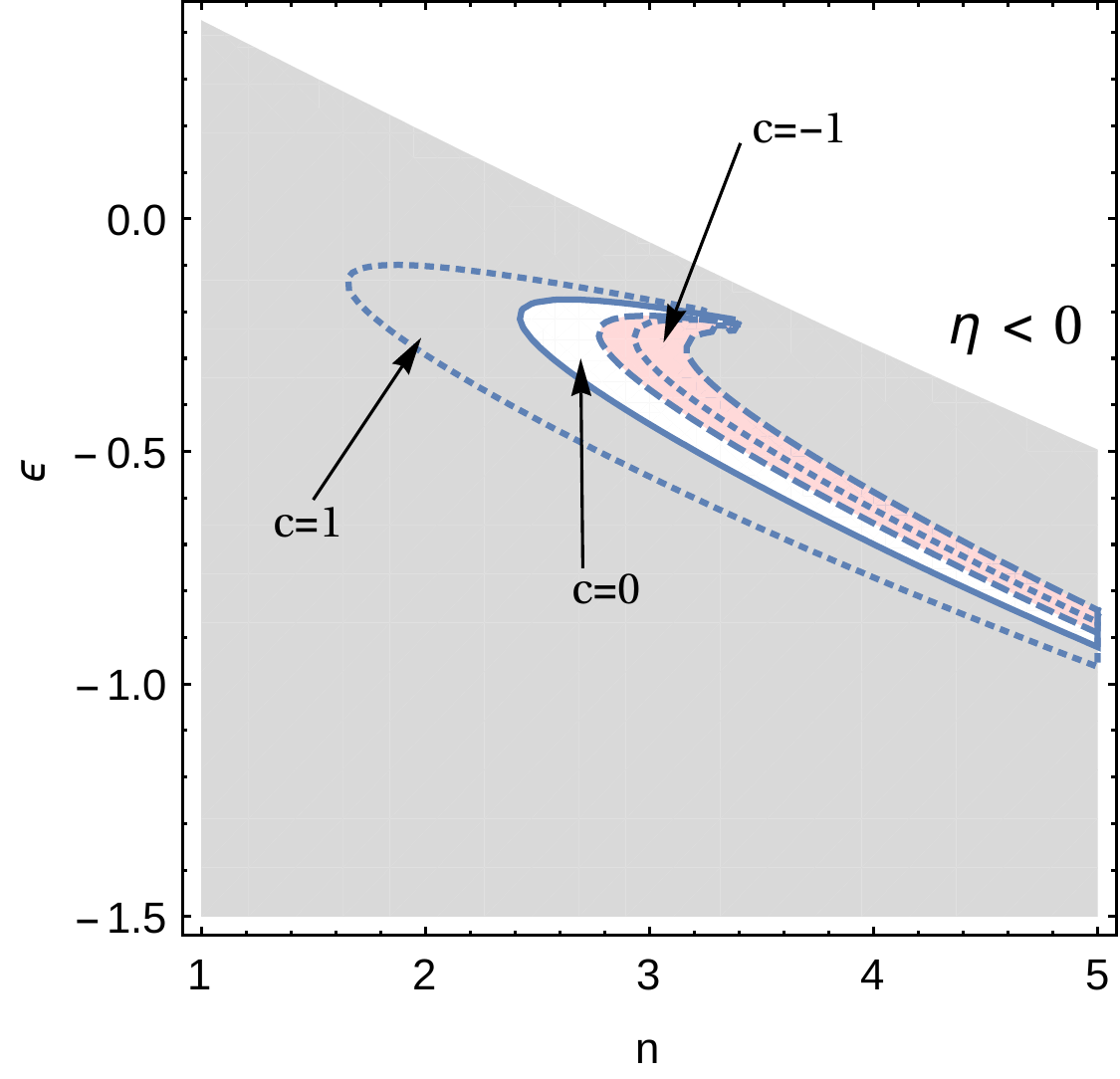}
\caption{\label{fig7}The allowed range for $n$ and $\epsilon$ for $\eta=0.01$ (left figure) and $\eta=-0.01$ (right figure), in the case $g(r)=r+\eta\, r^n e^{\epsilon r}$. Also, for each plot we have set $c_0=-1,0,+1$. In each figure, the gray area is the allowed region in which the first and the second derivative of $g(r)$ is positive.}
\end{figure}
In figure \eqref{fig7}‌ we have plotted the allowed range of the parameter space in the case $g(r)=r+\eta\, r^n e^{\epsilon r}$ assuming $\eta=\pm0.01$ and $c_0=0,\pm1$. For positive values of $\eta$ (left plot in figure \eqref{fig7}), the allowed parameter space is bounded. Also, one can see that smaller values for $\epsilon$‌ can be achieved by choosing positive values of $c_0$.

For negative values of $\eta$ (right plot in figure \eqref{fig7}) the parameter space of the model is unbounded and one can choose arbitrary large $n$. However, for accelerated expanding condition to be satisfied in this case, one should have $\epsilon<0$. Also, positive values of $c_0$ again extend the parameter space to smaller values of $n$ at given $\epsilon$ and $\eta$.

\section{Discussions and final remarks}\label{secDIS}
In this paper we have considered the $f(R)$ gravity in the context of mimetic dark matter theory. The mimetic dark matter theory introduces a very simple procedure to construct a conformally invariant gravitational action by expressing the physical metric in terms of a scalar field and a dynamical metric. Surprisingly, this procedure results in a degree of freedom which behaves like a dust and can be considered as a geometrical candidate of dark matter. The action is equivalent to a gravitational theory with a constraint which is added by a Lagrange multiplier. In this point of view, the Lagrange multiplier plays the role of the energy density of dark matter and the covariant derivative of the scalar field plays the role of its velocity. One of the interesting fact about the mimetic theory is that the equation of motion of the metric tensor has a general form
$$F_{\mu\nu}+F^\alpha_{~\alpha}\nabla_\mu\phi\nabla_\nu\phi=0.$$
Here $F_{\mu\nu}=0$ is the metric field equation in the non-mimetic gravity theory. The presence of scalar field results in the addition of $F^\mu_{~\mu}$ to the metric equation. In the Lagrange multiplier point of view, the trace of the metric equation is the Lagrange multiplier itself, which represents the energy density of the dark matter. 

The $f(R)$ gravity theory is equivalent to the special case of mimetic-$f(R)$ theory with vanishing $C_0$ (see equation \eqref{phi2}).  So, the mimetic-$f(R)$ theory is more general, in the sense that it yields more solutions and provides a wider range of parameters with respect to the cases of non-vanishing $C_0$. 

The mimetic-$f(R)$ theory is equivalent to a gravitational theory with two scalar fields. One of them is responsible for the dark matter sector of the universe and makes the theory conformally invariant. The other is a scalar field which is added to the action via Brans-Dicke interaction with $\omega=0$, and is responsible for the dark energy content of the universe and makes the universe accelerate.
The Dolgov-Kawasaki criterion can be found to be equivalent to the $f(R)$ gravity theory. This can be explained by the fact that the mimetic theory only adds a dust like scalar field which can not change the behavior of the expansion radically. However, as we have seen in this paper, the existence of this scalar field can change the parameter space of the theory.

The energy condition analysis of the theory shows that various forms of $f(R)$ can be used to obtain the accelerating universe. In the case $f(R)=R+\alpha R^n$, we have shown that generically, the allowed region, with respect to $n$, in which the accelerated expanding condition in fulfilled always lies in $n\in(n_2,n_3)$ where $n_2$ and $n_3$‌ corresponds to points in which the conditions (II) and (III) vanishes. Also, there is a special case for $\beta>0$ where the allowed range can extend to minus infinity.

 In the case $f(R)=\alpha\exp(\beta R)$, we have shown that in order that gravity remains attractive and the Dolgov-Kawasaki instability disappears, both the parameters $\beta$ and $\alpha$ should be positive. An interesting fact about this case is that the parameter $\beta$ always bounded but the parameter $\alpha$ can be extended to infinity. We have also considered a more general case $f(R)=R+\alpha R^n\exp(\beta R)$. We have seen that for both signs of the coupling $\alpha$, the $c_0=0$ case can produce a healthy accelerating universe.Also, we have shown in this case that the $c_0>0$ values gives more parameter range.

In summary, we have seen in this paper that the mimetic-$f(R)$ gravity generalizes the  $f(R)$ gravity theory which gives us more parameter space for having a healthy accelerated expanding phase. In this parameter space the Weak, Dominant and Null energy conditions remain valid, and the Dolgov-Kawasaki criterion is satisfied.

\end{document}